# GW170814: A Three-Detector Observation of Gravitational Waves from a Binary Black Hole Coalescence

B. P. Abbott et al.*

(LIGO Scientific Collaboration and Virgo Collaboration)

On August 14, 2017 at 10:30:43 UTC, the Advanced Virgo detector and the two Advanced LIGO detectors coherently observed a transient gravitational-wave signal produced by the coalescence of two stellar mass black holes, with a false-alarm rate of $\lesssim 1$ in 27 000 years. The signal was observed with a three-detector network matched-filter signal-to-noise ratio of 18. The inferred masses of the initial black holes are $30.5^{+5.7}_{-3.0} M_\odot$ and $25.3^{+2.8}_{-4.2} M_\odot$ (at the 90% credible level). The luminosity distance of the source is $540^{+130}_{-210}$ Mpc, corresponding to a redshift of $z = 0.11^{+0.03}_{-0.04}$. A network of three detectors improves the sky localization of the source, reducing the area of the 90% credible region from 1160 $\mathrm{deg}^2$ using only the two LIGO detectors to 60 $\mathrm{deg}^2$ using all three detectors. For the first time, we can test the nature of gravitational-wave polarizations from the antenna response of the LIGO-Virgo network, thus enabling a new class of phenomenological tests of gravity.

## I. INTRODUCTION

The era of gravitational-wave (GW) astronomy began with the detection of binary black hole (BBH) mergers, by the Advanced Laser Interferometer Gravitational-Wave Observatory (LIGO) detectors [1], during the first of the Advanced Detector Observation Runs. Three detections, GW150914 [2], GW151226 [3], and GW170104 [4], and a lower significance candidate, LVT151012 [5], have been announced so far. The Advanced Virgo detector [6] joined the second observation run on August 1, 2017.

On August 14, 2017, GWs from the coalescence of two black holes at a luminosity distance of $540^{+130}_{-210}$ Mpc, with masses of $30.5^{+5.7}_{-3.0} M_\odot$ and $25.3^{+2.8}_{-4.2} M_\odot$, were observed in all three detectors. The signal was first observed at the LIGO Livingston detector at 10:30:43 UTC, and at the LIGO Hanford and Virgo detectors with a delay of ~8 ms and ~14 ms, respectively.

The signal-to-noise ratio (SNR) time series, the time-frequency representation of the strain data, and the time series data of the three detectors together with the inferred GW waveform, are shown in Fig. 1. The different sensitivities and responses of the three detectors result in the GW producing different values of matched-filter SNR in each detector.

Three methods were used to assess the impact of the Virgo instrument on this detection. (a) Using the best fit waveform obtained from analysis of the LIGO detectors' data alone, we find that the probability, in 5000 s of data around the event, of a peak in SNR from Virgo data due to noise and as large as the one observed, within a time window determined by the maximum possible time of flight, is 0.3%. (b) A search for unmodeled GW transients demonstrates that adding Advanced Virgo improves the false-alarm rate by an order of magnitude over the two-detector network. (c) We compare the matched-filter marginal likelihood for a model with a coherent BBH signal in all three detectors to that for a model assuming pure Gaussian noise in Virgo and a BBH signal only in the LIGO detectors: the three detector BBH signal model is preferred with a Bayes factor of more than 1600.

Until Advanced Virgo became operational, typical GW position estimates were highly uncertain compared to the fields of view of most telescopes. The baseline formed by the two LIGO detectors allowed us to localize most mergers to roughly annular regions spanning hundreds to about a thousand square degrees at the 90% credible level [7–9]. Virgo adds additional independent baselines, which in cases such as GW170814 can reduce the positional uncertainty by an order of magnitude or more [8].

Tests of general relativity (GR) in the strong field regime have been performed with the signals from the BBH mergers detected by the LIGO interferometers [2–5,10]. In GR, GWs are characterized by two tensor (spin-2) polarizations only, whereas generic metric theories may allow up to six polarizations [11,12]. As the two LIGO instruments have similar orientations, little information about polarizations can be obtained using the LIGO detectors alone. With the addition of Advanced Virgo we can probe, for the first time, gravitational-wave polarizations geometrically by projecting the wave's amplitude

*Full author list given at the end of the Letter.

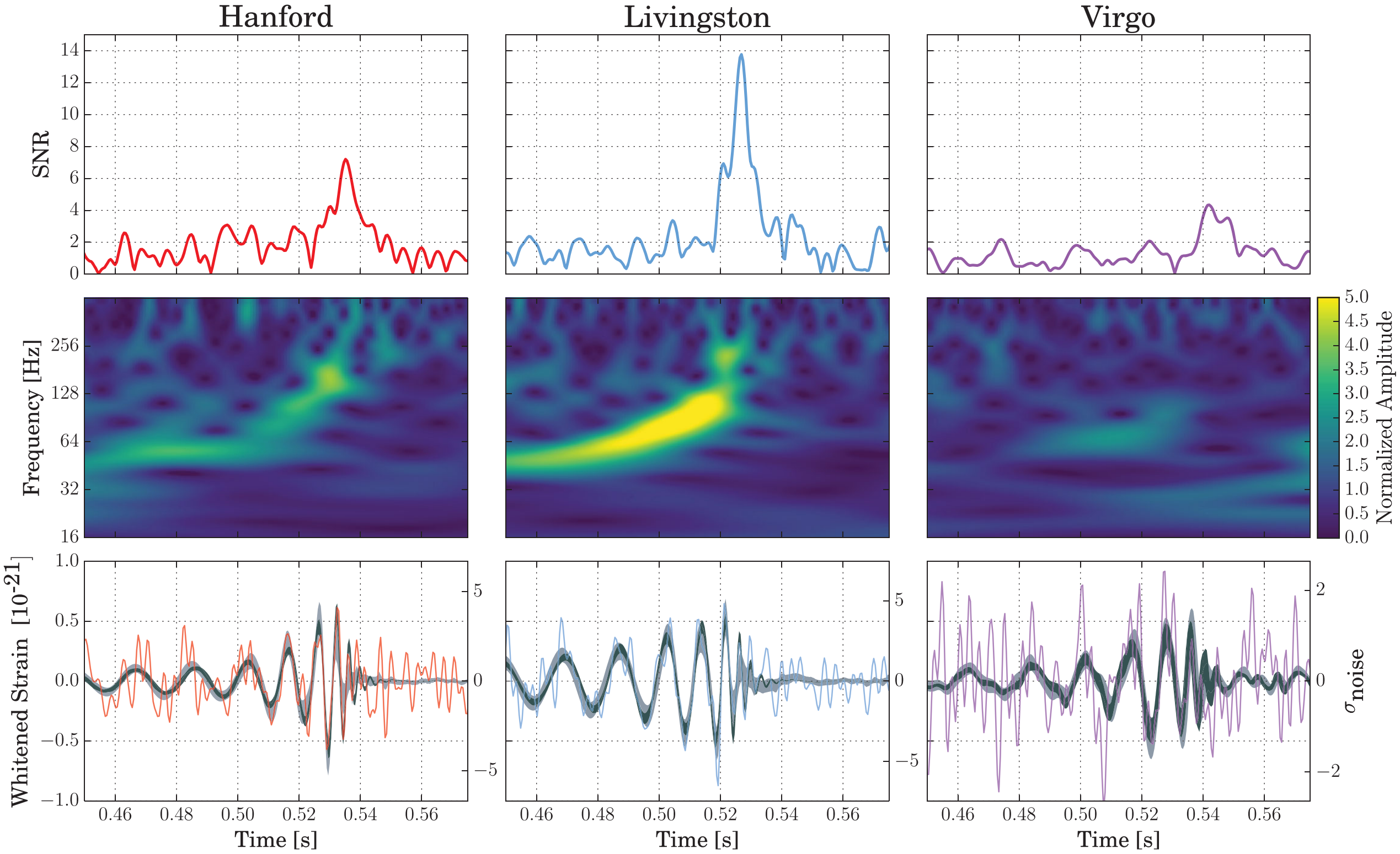

FIG. 1. The GW event GW170814 observed by LIGO Hanford, LIGO Livingston, and Virgo. Times are shown from August 14, 2017, 10:30:43 UTC. *Top row*: SNR time series produced in low latency and used by the low-latency localization pipeline on August 14, 2017. The time series were produced by time shifting the best-match template from the online analysis and computing the integrated SNR at each point in time. The single-detector SNRs in Hanford, Livingston, and Virgo are 7.3, 13.7, and 4.4, respectively. *Second row*: Time-frequency representation of the strain data around the time of GW170814. *Bottom row*: Time-domain detector data (in color), and 90% confidence intervals for waveforms reconstructed from a morphology-independent wavelet analysis [13] (light gray) and BBH models described in Sec. V (dark gray), whitened by each instrument's noise amplitude spectral density between 20 Hz and 1024 Hz. For this figure the data were also low passed with a 380 Hz cutoff to eliminate out-of-band noise. The whitening emphasizes different frequency bands for each detector, which is why the reconstructed waveform amplitude evolution looks different in each column. The left ordinate axes are normalized such that the physical strain of the wave form is accurate at 130 Hz. The right ordinate axes are in units of whitened strain, divided by the square root of the effective bandwidth (360 Hz), resulting in units of noise standard deviations.

onto the three detectors. As an illustration, we perform a test comparing the tensor-only mode with scalar-only and vector-only modes. We find that purely tensor polarization is strongly favored over purely scalar or vector polarizations. With this, and additional tests, we find that GW170814 is consistent with GR.

## II. DETECTORS

LIGO operates two 4 km long detectors in the U.S., one in Livingston, LA and one in Hanford, WA [14], while Virgo consists of a single 3 km long detector near Pisa, Italy [15]. Together with GEO600 located near Hanover, Germany [16], several science runs of the initial-era gravitational-wave network were conducted through 2011. LIGO stopped observing in 2010 for the Advanced LIGO upgrade [1]. The Advanced LIGO detectors have been operational since 2015 [17]. They underwent a series of upgrades between the first and second observation runs [4], and began observing again in November 2016.

Virgo stopped observing in 2011 for the Advanced Virgo upgrade, during which many parts of the detector were replaced or improved [6]. Among the main changes are an increase of the finesse of the arm cavities, the use of heavier test mass mirrors that have lower absorption and better surface quality [18]. To reduce the impact of the coating thermal noise [19], the size of the beam in the central part of the detector was doubled, which required modifications of the vacuum system and the input-output optics [20,21]. The recycling cavities are kept marginally stable as in the initial Virgo configuration. The optical benches supporting the main readout photodiodes have been suspended and put under vacuum to reduce the impact of scattered light and acoustic noise. Cryogenic traps have been installed to improve the vacuum level. The vibration isolation and suspension system, already compliant with the Advanced Virgo requirement [22,23], has been further improved to allow for a more robust control of the last-stage pendulum and the accommodation of baffles to mitigate the effect of scattered light. The test mass mirrors are currently suspended with metallic wires. Following one year of commissioning, Advanced Virgo joined LIGO in August 2017 for the last month of the second observation run.

For Virgo, the noises that are currently limiting the sensitivity at low frequencies are thermal noise of the test mass suspension wires, control noise, the 50 Hz mains line and harmonics, and scattered light driven by seismic noise. At high frequencies, the largest contribution comes from shot noise of the main interferometer beam, with smaller contributions coming from scattered light, and shot noise of a secondary beam used to control the laser frequency. The noise sources that limit LIGO's sensitivity are described in [24] and [25]. For both LIGO and Virgo, commissioning will continue to reach their ultimate designed sensitivities [26].

Several noise sources that are linearly coupled to the GW data channel can be subtracted in postprocessing, using auxiliary sensors (e.g., photodiodes monitoring beam motion) and coupling transfer functions calculated via optimal Wiener filters. This technique was used in the initial detector era [27–29]. For LIGO, we remove calibration lines, power mains and harmonics, the effect of some length and angular controls, and the effect of laser beam motion. This noise removal can improve the sensitivity of the LIGO detectors by approximately 20% [30]. For Virgo, we remove the effect of some length controls, and the laser frequency stabilization control. The search pipelines described in Sec. III use the calibrated strain data which were produced in low latency and which have not undergone postprocessing noise subtraction. They also use data quality flags which were produced offline. The source properties, however, described in Sec. V, are inferred using the postprocessing noise-subtracted data. Figure 2 shows the sensitivity of the Advanced LIGO–Advanced Virgo network around the time of GW170814, after the postprocessing removal of several noise sources.

Detection validation procedures at LIGO [2,31], and checks performed at Virgo found no evidence that instrumental or environmental disturbances could account for GW170814. Tests quantifying the detectors' susceptibility to external environmental disturbances, such as electromagnetic fields [32], indicated that any disturbance strong enough to account for the signal would be clearly detected by the array of environmental sensors. None of the environmental sensors recorded any disturbances consistent with a signal that evolved in time and frequency like GW170814. A noise transient with a central frequency around 50 Hz occurs in the Virgo detector 50 ms after GW170814. This falls outside the window expected due to the light travel time between the detectors, and has, therefore, no effect on the interpretation of the GW signal.

LIGO is calibrated by inducing test-mass motion using photon pressure from modulated auxiliary lasers [33,34], and Virgo is calibrated using electromagnetic actuators [35,36]. Frequency-dependent calibration uncertainties are determined for both LIGO detectors for GW170814 using the method in [37], and used for estimation of the properties of this event; the maximum 1-$\sigma$ uncertainty for the strain data in the frequency range 20–1024 Hz is 7% in amplitude and 4° in phase. The maximum 1-$\sigma$ uncertainties for Virgo are 8% in amplitude and 3° in phase over the same frequency range. The estimation of properties of GW170814 use these maximum values for the Virgo uncertainty over the whole frequency range. Uncertainties in the time stamping of the data are 10 $\mu$s for LIGO and 20 $\mu$s for Virgo, which does not limit the sky localization.

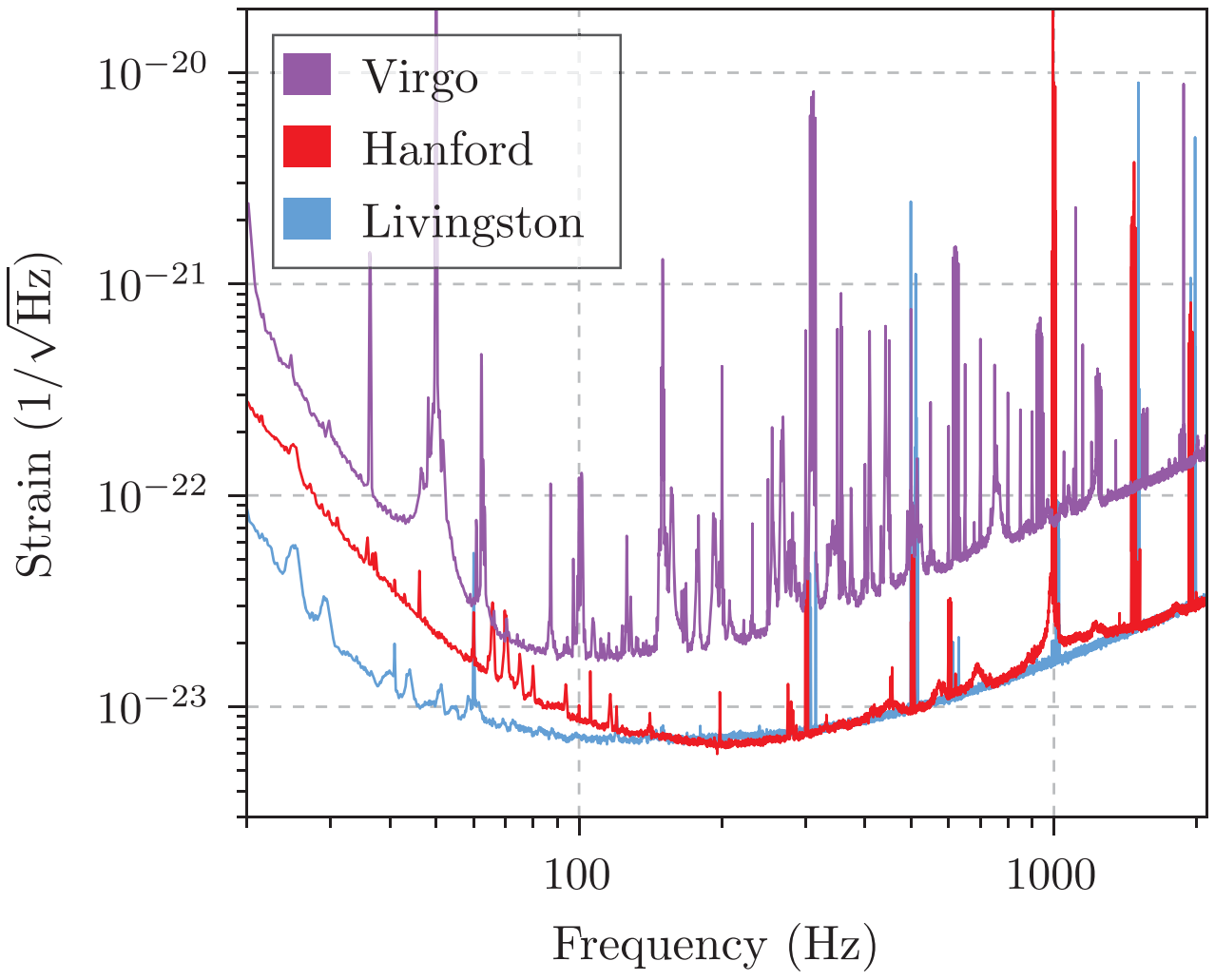

FIG. 2. Amplitude spectral density of strain sensitivity of the Advanced LIGO–Advanced Virgo network, estimated using 4096 s of data around the time of GW170814. Here, several known linearly coupled noise sources have been removed from the data.

## III. SEARCHES

GW170814 was first identified with high confidence ~30 s after its arrival by two independent low-latency matched-filter pipelines [38–44] that filter the data against a collection of approximate gravitational-wave templates [45–53], triggering an alert that was shared with partners for electromagnetic follow-up [54].

The significance estimates for this event were found by the two matched-filter pipelines, and a fully coherent unmodeled search pipeline [55], analyzing 5.9 days of coincident strain data from the Advanced LIGO detectors spanning August 13, 2017 to August 21, 2017. The matched-filter pipelines do not currently use data from Virgo for significance estimates. Coherent searches, however, use the Virgo data to improve significance estimates.

The analysis was performed over the same source parameter space as the GW170104 matched-filter analysis [4] and with additional data quality information unavailable in low latency [5,31], although the noise-subtracted data described in Sec. II were not used. Both pipelines identified GW170814 with a Hanford-Livingston network SNR of 15, with ranking statistic values from the two pipelines corresponding to a false-alarm rate of 1 in 140 000 years in one search [38,39] and 1 in 27 000 years in the other search [40–44,56], clearly identifying GW170814 as a GW signal. The difference in significance is due to the different techniques used to rank candidate events and measure the noise background in these searches; however, both report a highly significant event.

The significance of GW170814 was confirmed on the full network of three detectors by an independent coherent analysis that targets generic gravitational-wave transients with increasing frequency over time [55]. This more generic search reports a false-alarm rate $< 1$ in 5900 years. By comparison, when we limit this analysis to the two LIGO detectors only, the false-alarm rate is approximately 1 in 300 years; the use of the data from Virgo improves significance by more than an order of magnitude. Moreover, this independent approach recovers waveforms and SNRs at the three detectors which are compatible with respect to the coherent analyses used to infer source properties (see Sec. V).

## IV. LOCALIZATION

Some compact object mergers are thought to produce not just GWs but also broadband electromagnetic emission. LIGO and Virgo have been distributing low-latency alerts and localizations of GW events to a consortium now consisting of ground- and space-based facilities who are searching for gamma-ray, x-ray, optical, near-infrared, radio, and neutrino counterparts [57–59].

For the purpose of position reconstruction, the LIGO-Virgo GW detector network can be thought of as a phased array of antennas. Any single detector provides only minimal position information, its slowly varying antenna pattern favoring two broad regions perpendicular to the plane of the detectors' arms [60,61]. However, with a network of detectors, sky position can be inferred by triangulation employing the time differences [62,63], phase differences, and amplitude ratios on arrival at the sites [64].

An initial rapid localization was performed by coherent triangulation of the matched-filter estimates of the times, amplitudes, and phases on arrival [65]. The localization was then progressively refined by full coherent Bayesian parameter estimation [66], using more sophisticated waveform models and treatment of calibration systematics, as described in the next section.

The localization of GW170814 is shown in Fig. 3. For the rapid localization from Hanford and Livingston, the 90% credible area on the sky is 1160 $\mathrm{deg}^2$ and shrinks to 100 $\mathrm{deg}^2$ when including Virgo data. The full parameter estimation further constrains the position to a 90% credible area of 60 $\mathrm{deg}^2$ centered at the maximum *a posteriori* position of right ascension $\mathrm{RA} = 03^{\mathrm{h}}11^{\mathrm{m}}$ and declination $\mathrm{dec} = -44°57^{\mathrm{m}}$ (J2000). The shift between the rapid localization and the full parameter estimation is partly due to the noise removal and final detector calibration, described in the previous section, that was applied for the full parameter estimation but not the rapid localization.

Incorporating Virgo data also reduces the luminosity distance uncertainty from $570^{+300}_{-230}$ Mpc (rapid localization) to $540^{+130}_{-210}$ Mpc (full parameter estimation). As with the previous paragraph, the three-dimensional credible volume and number of possible host galaxies also decreases by an order of magnitude [67–69], from $71 \times 10^6\ \mathrm{Mpc}^3$, to $3.4 \times 10^6\ \mathrm{Mpc}^3$, to $2.1 \times 10^6\ \mathrm{Mpc}^3$.

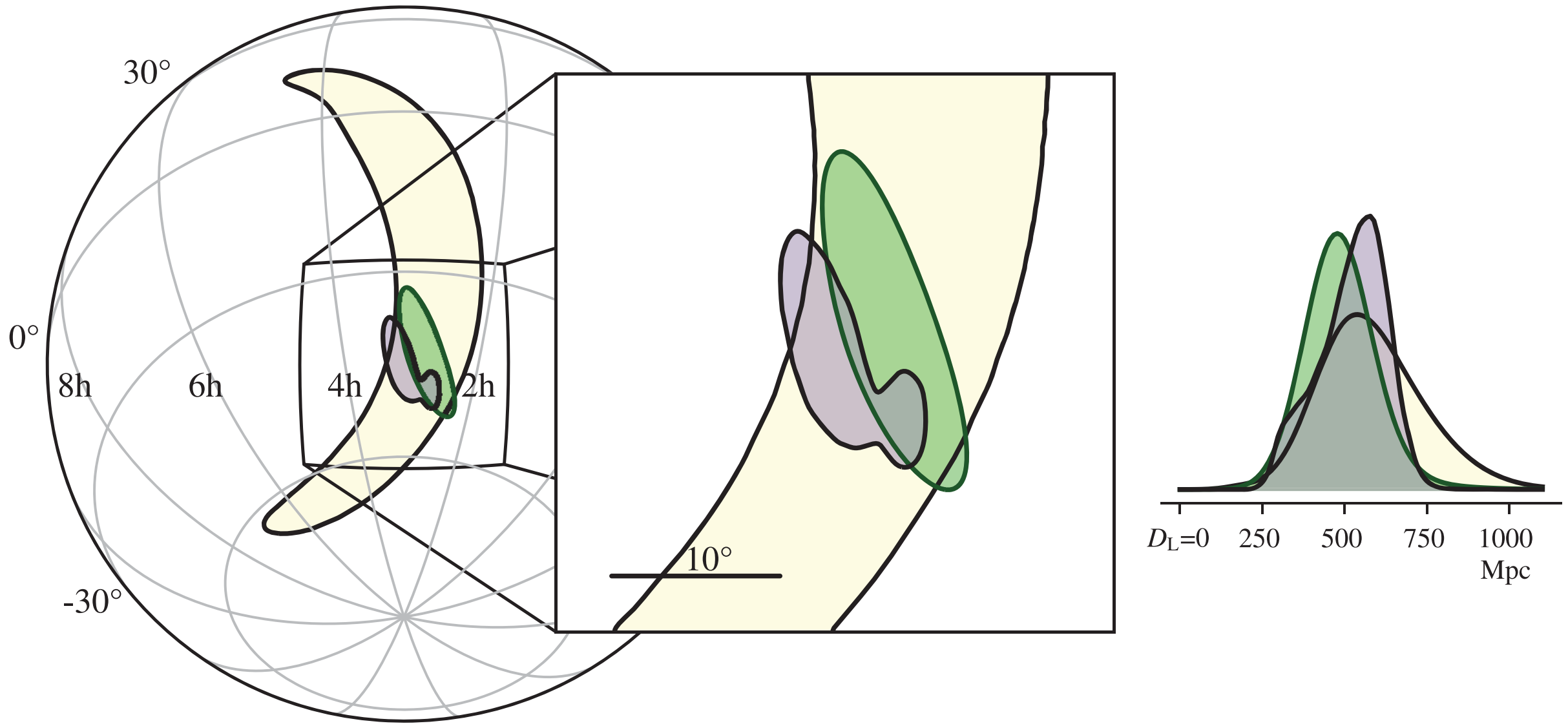

FIG. 3. Localization of GW170814. The rapid localization using data from the two LIGO sites is shown in yellow, with the inclusion of data from Virgo shown in green. The full Bayesian localization is shown in purple. The contours represent the 90% credible regions. The left panel is an orthographic projection and the inset in the center is a gnomonic projection; both are in equatorial coordinates. The inset on the right shows the posterior probability distribution for the luminosity distance, marginalized over the whole sky.

Follow-up observations of GW170814 were conducted by 25 facilities in neutrinos [70–72], gamma rays [73–81], x rays [82–85], and in optical and near infrared [86–98]. No counterpart has been reported so far.

## V. SOURCE PROPERTIES

The parameters of the source are inferred through a coherent Bayesian analysis [66,99] of offline noise-subtracted data for the LIGO and Virgo detectors using two independently developed waveform models.

Both of these waveform models are calibrated to partially overlapping sets of numerical-relativity simulations of binary black hole coalescences, following from the initial breakthroughs reported in [100–102]. One model includes the full two-spin inspiral dynamics in the absence of precession [53,103–109,109], whereas the other model includes an effective treatment of the spin-precession dynamics through a rotation of an originally nonprecessing model [110–113]. Previous studies [114–117] have investigated the effect of systematic waveform modeling in sources we believe to be similar to GW170814, albeit for a different detector network configuration. Based on these analogous investigations, and due to the brevity and amplitude of GW170814, we expect systematic biases to be significantly smaller than the statistical error reported in this work.

In addition to a waveform model, the coherent Bayesian analysis also incorporates the detectors' noise power spectral densities at the time of the event [13,118] and marginalizes over the calibration uncertainties described in Sec. II, as in [4,5,119]. We note that the likelihood used in our analyses assumes that the noise in the detectors is Gaussian in the 4 s window around the event. While some non-Gaussian and nonstationary features exist in the data, initial investigations suggest that the non-Gaussian features in the data do not significantly impact the reported parameters, but we defer a detailed study of these effects to future work. The coherent Bayesian analysis recovers the maximum matched-filter SNR across the LIGO-Virgo network of 18.3 [66,120], with individual detector matched-filter SNRs of 9.7, 14.8, and 4.8 in LIGO Hanford, LIGO Livingston, and Virgo, respectively.

Table I shows source parameters for GW170814, where we quote the median value and the symmetric 90% credible intervals. The final mass (or equivalently the energy radiated), final spin, and peak luminosity are computed from averages of fits to numerical relativity simulations [121–125]. The reported uncertainties account for both statistical and systematic uncertainties from averaging over the two waveform models used. An independent calculation using direct comparison to numerical relativity gives consistent parameters [114].

The inferred posterior distributions for the two black hole masses $m_1$ and $m_2$ are shown in Fig. 4. GW170814 allows for measurements of comparable accuracy of the total binary mass $M = m_1 + m_2$, which is primarily governed by the merger and ringdown, and the chirp mass $\mathcal{M} = (m_1 m_2)^{3/5}/M^{1/5}$, determined by the binary inspiral [64,131–137], similarly to both GW150914 [99] and GW170104 [4].

TABLE I. Source parameters for GW170814: median values with 90% credible intervals. We quote source-frame masses; to convert to the detector frame, multiply by $(1+z)$ [126,127]. The redshift assumes a flat cosmology with Hubble parameter $H_0 = 67.9\ \mathrm{km\,s^{-1}\,Mpc^{-1}}$ and matter density parameter $\Omega_m = 0.3065$ [128].

| | |
|---|---|
| Primary black hole mass $m_1$ | $30.5^{+5.7}_{-3.0} M_\odot$ |
| Secondary black hole mass $m_2$ | $25.3^{+2.8}_{-4.2} M_\odot$ |
| Chirp mass $\mathcal{M}$ | $24.1^{+1.4}_{-1.1} M_\odot$ |
| Total mass $M$ | $55.9^{+3.4}_{-2.7} M_\odot$ |
| Final black hole mass $M_f$ | $53.2^{+3.2}_{-2.5} M_\odot$ |
| Radiated energy $E_{\rm rad}$ | $2.7^{+0.4}_{-0.3} M_\odot\ c^2$ |
| Peak luminosity $\ell_{\rm peak}$ | $3.7^{+0.5}_{-0.5} \times 10^{56}\ \mathrm{erg\,s^{-1}}$ |
| Effective inspiral spin parameter $\chi_{\rm eff}$ | $0.06^{+0.12}_{-0.12}$ |
| Final black hole spin $a_f$ | $0.70^{+0.07}_{-0.05}$ |
| Luminosity distance $D_L$ | $540^{+130}_{-210}$ Mpc |
| Source redshift $z$ | $0.11^{+0.03}_{-0.04}$ |

The orbital evolution is dominated by the black hole masses and the components of their spins $\mathbf{S}_{1,2}$ perpendicular to the orbital plane, and other spin components affect the GW signal on a subdominant level. The dominant spin effects are represented through the effective inspiral spin parameter $\chi_{\rm eff} = (m_1 a_1 \cos\theta_{LS_1} + m_2 a_2 \cos\theta_{LS_2})/M$ which is approximately conserved throughout the evolution of the binary orbit [138–141]. Here, $\theta_{LS_i}$ is the angle between the black hole spin $\mathbf{S}_i$ and the Newtonian orbital angular momentum $\mathbf{L}$ for both the primary $(i=1)$ and secondary $(i=2)$ black holes, and $a_i = |c\mathbf{S}_i/Gm_i^2|$ is the dimensionless spin magnitude of the initial $(i=1,2)$ and final $(i=f)$ black holes. For $a_{1,2}$, this analysis assumed a uniform prior distribution between 0 and 0.99, with no restrictions on the spin orientations. As with GW150914 and GW170104, $\chi_{\rm eff}$ is consistent with having a arbitrarily small value [4,5]. The spin components orthogonal to $\mathbf{L}$ are interesting, as they lead to a precession of the binary orbit [142,143] and are here quantified by the effective precession spin parameter $\chi_p$ [112,143]. As for previous events [4,5,116,130], the $\chi_p$ posterior distribution is dominated by assumptions about the prior, as shown in Fig. 4. Given these assumptions, as well as statistical and systematic uncertainties, we cannot draw further robust conclusions about the transverse components of the spin. The event, GW170814, is consistent with the population of BBHs, physical parameters, and merger rate reported in previous BBH papers [5,144,145].

The accuracy with which parameters influencing the phase evolution of the observed GW, the black hole masses and spins, can be measured is determined by the network SNR. For GW170814 this is dominated by the two LIGO detectors. The inclusion of Virgo data into the coherent analysis significantly improves the inference of parameters describing the binary's position relative to the Earth, as shown in Fig. 3, since those parameters are predominantly determined by the relative amplitudes and arrival times observed in the detector network [67,146,147]. Because of the inferred orientation of the binary, we do not see a significant improvement in parameters such as inclination and polarization angle for GW170814.

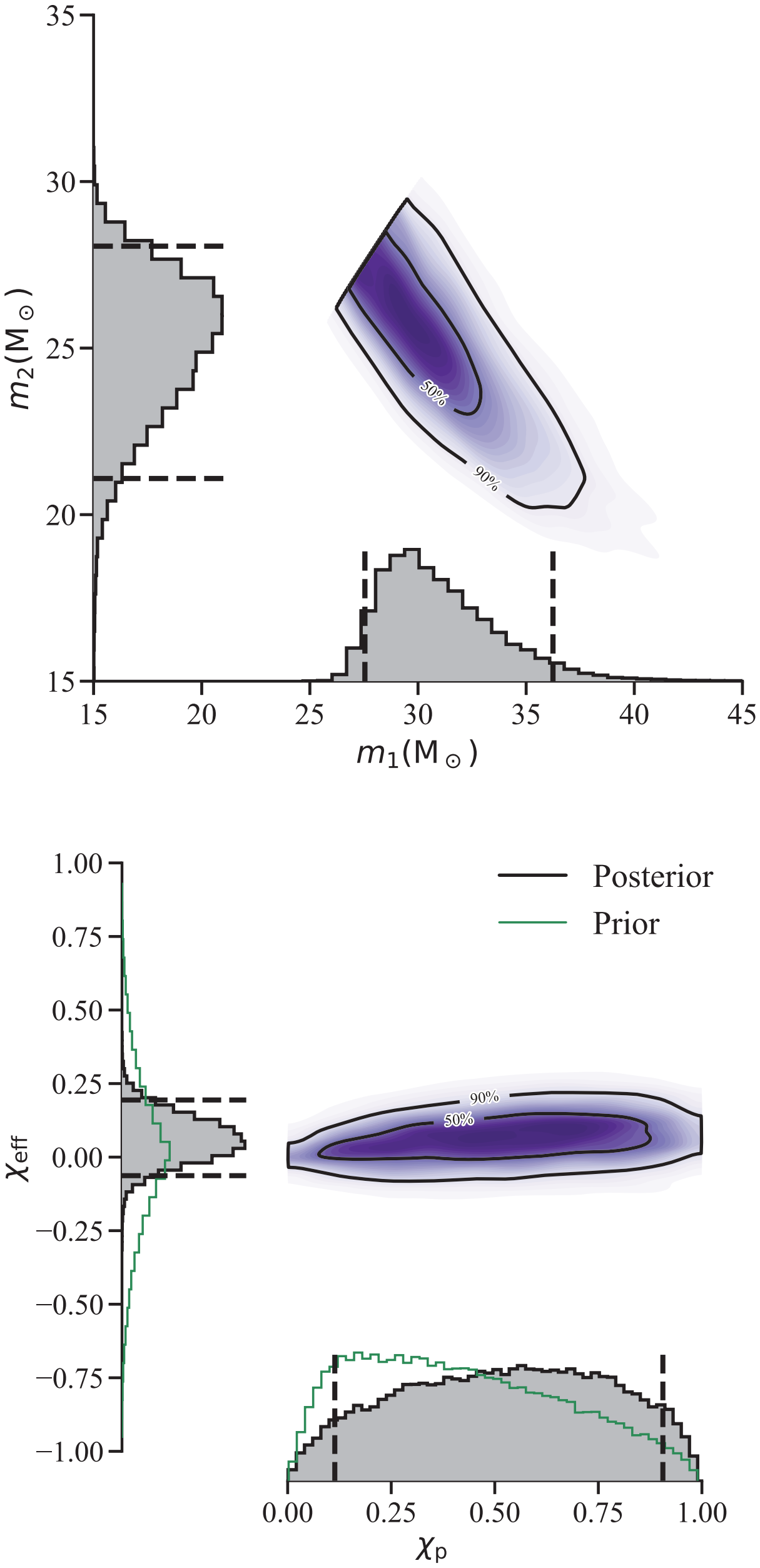

FIG. 4. Posterior probability density for the source-frame masses $m_1$ and $m_2$ (*top*) and the effective inspiral and precession spin parameters, $\chi_{eff}$ and $\chi_p$ (*bottom*) measured at a gravitational-wave frequency of 20 Hz, well before the merger. The dashed lines mark the 90% credible interval for the one-dimensional marginalized distributions. The two-dimensional plots show the contours of the 50% and 90% credible regions plotted over a color-coded posterior density function. For GW170814, both $\chi_{eff}$ and $\chi_p$ are influenced by their respective prior distributions, shown in green. While the GW observation provides additional constraints for the $\chi_{eff}$ posterior, there is only a marginal information gain for $\chi_p$. (Kullback–Leibler divergence between the prior and posterior distribution of 0.08 nat [4,129,130].)

## VI. TESTS OF GENERAL RELATIVITY

To determine the consistency of the signal with GR, we allowed the post-Newtonian (PN) and additional coefficients describing the waveform to deviate from their nominal values [148–150], as was done for previous detections [2–5,10]. In addition to previously tested coefficients, these analyses were expanded to also explicitly consider phase contributions at effective −1PN order, i.e., with a frequency dependence of $f^{-7/3}$. Additionally, as in [2–4], we check that the inspiral and merger-ringdown regimes are mutually consistent, and check for possible deviations from GR in the propagation of GWs due to a massive graviton and/or Lorentz invariance violation. Preliminary results of all these tests show no evidence for disagreement with the predictions of GR; detailed investigations are still ongoing, and full results will be presented at a later date.

## VII. GRAVITATIONAL-WAVE POLARIZATIONS

One of the key predictions of GR is that metric perturbations possess two tensor degrees of freedom [151,152]. These two are only a subset of the six independent modes allowed by generic metric theories of gravity, which may in principle predict any combination of tensor (spin-2), vector (spin-1), or scalar (spin-0) polarizations [11,12]. While it may be that any generic theory of gravity will be composed of a potential mixture of polarization modes, an investigation of this type is beyond the scope of this Letter. However, a simplified first investigation that serves to illustrate the potential power of this new phenomenological test of gravity is to consider models where the polarization states are pure tensor, pure vector, or pure scalar only.

So far, some evidence that GWs are described by the tensor (spin-2) metric perturbations of GR has been obtained from measurements of the rate of orbital decay of binary pulsars, in the context of specific beyond-GR theories (see, e.g., [153,154] or [155,156] for reviews), and from the rapidly changing GW phase of BBH mergers observed by LIGO, in the framework of parametrized models [2,4,10]. The addition of Advanced Virgo provides us with another, more compelling, way of probing the nature of polarizations by studying GW geometry directly through the projection of the metric perturbation onto our detector network [157–159].

The coherent Bayesian analysis described in Sec. V is repeated after replacing the standard tensor antenna response functions with those appropriate for scalar or vector polarizations [160]. In our analysis, we are interested in the geometric projection of the GW onto the detector network; therefore, the details of the phase model itself are less relevant as long it is a faithful representation of the fit to the data in Fig 1. Hence, we assume a GR phase model. We find Bayes' factors of more than 200 and 1000 in favor of the purely tensor polarization against purely vector and purely scalar, respectively. We also find that, as expected, the reconstructed sky location and distance change significantly depending on the polarization content of the source, with nonoverlapping 90% credible regions for tensor, vector, and scalar. The inferred masses and spins are always the same, because that information is encoded in the signal phasing. These are only the simplest possible phenomenological models, but a more intensive study involving mixed-polarization states, using both matched-filter and generic GW transient models, is currently under way. Similar tests were inconclusive for previous events [10] because the two LIGO detectors are very nearly coaligned, and record the same combination of polarizations.

## VIII. CONCLUSION

On August 1, 2017, Advanced Virgo joined the two Advanced LIGO detectors in the second Advanced Detector Observation Runs. On August 14, 2017, a GW signal coming from the merger of two stellar mass black holes was observed with the Virgo and LIGO detectors. The three-detector detection of GW170814 allowed for a significant reduction in the search volume for the source. The black hole characteristics of GW170814 are similar to GW150914 and GW170104, and are found to be consistent with the astrophysical population and merger rate determined with previous detections. The addition of Virgo has allowed us to probe the polarization content of the signal for the first time; we find that the data strongly favor pure tensor polarization of gravitational waves, over pure scalar or pure vector polarizations. Data for this event are available at the LIGO Open Science Center [161].

The authors gratefully acknowledge the support of the United States National Science Foundation (NSF) for the construction and operation of the LIGO Laboratory and Advanced LIGO as well as the Science and Technology Facilities Council (STFC) of the United Kingdom, the Max-Planck-Society (MPS), and the State of Niedersachsen, Germany, for support of the construction of Advanced LIGO and construction and operation of the GEO600 detector. Additional support for Advanced LIGO was provided by the Australian Research Council. The authors gratefully acknowledge the Italian Istituto Nazionale di Fisica Nucleare (INFN), the French Centre National de la Recherche Scientifique (CNRS), and the Foundation for Fundamental Research on Matter supported by the Netherlands Organisation for Scientific Research, for the construction and operation of the Virgo detector and the creation and support of the EGO consortium. The authors also gratefully acknowledge research support from these agencies as well as by the Council of Scientific and Industrial Research of India, the Department of Science and Technology, India, the Science & Engineering Research Board (SERB), India, the Ministry of Human Resource Development, India, the Spanish Agencia Estatal de Investigación, the Vicepresidència i Conselleria d'Innovació, Recerca i Turisme and the Conselleria d'Educació i Universitat del Govern de les Illes Balears, the Conselleria d'Educació, Investigació, Cultura i Esport de la Generalitat Valenciana, the National Science Centre of Poland, the Swiss National Science Foundation (SNSF), the Russian Foundation for Basic Research, the Russian Science Foundation, the European Commission, the European Regional Development Funds (ERDF), the Royal Society, the Scottish Funding Council, the Scottish Universities Physics Alliance, the Hungarian Scientific Research Fund (OTKA), the Lyon Institute of Origins (LIO), the National Research, Development and Innovation Office Hungary (NKFI), the National Research Foundation of Korea, Industry Canada and the Province of Ontario through the Ministry of Economic Development and Innovation, the Natural Sciences and Engineering Research Council Canada, the Canadian Institute for Advanced Research, the Brazilian Ministry of Science, Technology, Innovations, and Communications, the International Center for Theoretical Physics South American Institute for Fundamental Research (ICTP-SAIFR), the Research Grants Council of Hong Kong, the National Natural Science Foundation of China (NSFC), the Leverhulme Trust, the Research Corporation, the Ministry of Science and Technology (MOST), Taiwan and the Kavli Foundation. The authors gratefully acknowledge the support of the NSF, STFC, MPS, INFN, CNRS, and the State of Niedersachsen, Germany, for provision of computational resources.

B. P. Abbott,[1] R. Abbott,[1] T. D. Abbott,[2] F. Acernese,[3,4] K. Ackley,[5,6] C. Adams,[7] T. Adams,[8] P. Addesso,[9] R. X. Adhikari,[1] V. B. Adya,[10] C. Affeldt,[10] M. Afrough,[11] B. Agarwal,[12] M. Agathos,[13] K. Agatsuma,[14] N. Aggarwal,[15] O. D. Aguiar,[16] L. Aiello,[17,18] A. Ain,[19] P. Ajith,[20] B. Allen,[10,21,22] G. Allen,[12] A. Allocca,[23,24] P. A. Altin,[25] A. Amato,[26] A. Ananyeva,[1] S. B. Anderson,[1] W. G. Anderson,[21] S. V. Angelova,[27] S. Antier,[28] S. Appert,[1] K. Arai,[1] M. C. Araya,[1] J. S. Areeda,[29] N. Arnaud,[28,30] K. G. Arun,[31] S. Ascenzi,[32,33] G. Ashton,[10] M. Ast,[34] S. M. Aston,[7] P. Astone,[35] D. V. Atallah,[36] P. Aufmuth,[22] C. Aulbert,[10] K. AultONeal,[37] C. Austin,[2] A. Avila-Alvarez,[29] S. Babak,[38] P. Bacon,[39] M. K. M. Bader,[14] S. Bae,[40] P. T. Baker,[41] F. Baldaccini,[42,43] G. Ballardin,[30] S. W. Ballmer,[44] S. Banagiri,[45] J. C. Barayoga,[1] S. E. Barclay,[46] B. C. Barish,[1] D. Barker,[47] K. Barkett,[48] F. Barone,[3,4] B. Barr,[46] L. Barsotti,[15] M. Barsuglia,[39] D. Barta,[49] S. D. Barthelmy,[50] J. Bartlett,[47] I. Bartos,[51,5] R. Bassiri,[52] A. Basti,[23,24] J. C. Batch,[47] M. Bawaj,[53,43] J. C. Bayley,[46] M. Bazzan,[54,55] B. Bécsy,[56] C. Beer,[10] M. Bejger,[57] I. Belahcene,[28] A. S. Bell,[46] B. K. Berger,[1] G. Bergmann,[10] J. J. Bero,[58] C. P. L. Berry,[59] D. Bersanetti,[60] A. Bertolini,[14] J. Betzwieser,[7] S. Bhagwat,[44] R. Bhandare,[61] I. A. Bilenko,[62] G. Billingsley,[1] C. R. Billman,[5] J. Birch,[7] R. Birney,[63] O. Birnholtz,[10] S. Biscans,[1,15] S. Biscoveanu,[64,6] A. Bisht,[22] M. Bitossi,[30,24] C. Biwer,[44] M. A. Bizouard,[28] J. K. Blackburn,[1] J. Blackman,[48] C. D. Blair,[1,65] D. G. Blair,[65] R. M. Blair,[47] S. Bloemen,[66] O. Bock,[10] N. Bode,[10] M. Boer,[67] G. Bogaert,[67] A. Bohe,[38] F. Bondu,[68] E. Bonilla,[52] R. Bonnand,[8] B. A. Boom,[14] R. Bork,[1] V. Boschi,[30,24] S. Bose,[69,19] K. Bossie,[7] Y. Bouffanais,[39] A. Bozzi,[30] C. Bradaschia,[24] P. R. Brady,[21] M. Branchesi,[17,18] J. E. Brau,[70] T. Briant,[71] A. Brillet,[67] M. Brinkmann,[10] V. Brisson,[28] P. Brockill,[21] J. E. Broida,[72] A. F. Brooks,[1] D. A. Brown,[44] D. D. Brown,[73] S. Brunett,[1] C. C. Buchanan,[2] A. Buikema,[15] T. Bulik,[74] H. J. Bulten,[75,14] A. Buonanno,[38,76] D. Buskulic,[8] C. Buy,[39] R. L. Byer,[52] M. Cabero,[10] L. Cadonati,[77] G. Cagnoli,[26,78] C. Cahillane,[1] J. Calderón Bustillo,[77] T. A. Callister,[1] E. Calloni,[79,4] J. B. Camp,[50] M. Canepa,[80,60] P. Canizares,[66] K. C. Cannon,[81] H. Cao,[73] J. Cao,[82] C. D. Capano,[10] E. Capocasa,[39] F. Carbognani,[30] S. Caride,[83] M. F. Carney,[84] J. Casanueva Diaz,[28] C. Casentini,[32,33]

S. Caudill,[21,14] M. Cavaglià,[11] F. Cavalier,[28] R. Cavalieri,[30] G. Cella,[24] C. B. Cepeda,[1] P. Cerdá-Durán,[85] G. Cerretani,[23,24] E. Cesarini,[86,33] S. J. Chamberlin,[64] M. Chan,[46] S. Chao,[87] P. Charlton,[88] E. Chase,[89] E. Chassande-Mottin,[39] D. Chatterjee,[21] K. Chatziioannou,[90] B. D. Cheeseboro,[41] H. Y. Chen,[91] X. Chen,[65] Y. Chen,[48] H.-P. Cheng,[5] H. Chia,[5] A. Chincarini,[60] A. Chiummo,[30] T. Chmiel,[84] H. S. Cho,[92] M. Cho,[76] J. H. Chow,[25] N. Christensen,[72,67] Q. Chu,[65] A. J. K. Chua,[13] S. Chua,[71] A. K. W. Chung,[93] S. Chung,[65] G. Ciani,[5,54,55] R. Ciolfi,[94,95] C. E. Cirelli,[52] A. Cirone,[80,60] F. Clara,[47] J. A. Clark,[77] P. Clearwater,[96] F. Cleva,[67] C. Cocchieri,[11] E. Coccia,[17,18] P.-F. Cohadon,[71] D. Cohen,[28] A. Colla,[97,35] C. G. Collette,[98] L. R. Cominsky,[99] M. Constancio Jr.,[16] L. Conti,[55] S. J. Cooper,[59] P. Corban,[7] T. R. Corbitt,[2] I. Cordero-Carrión,[100] K. R. Corley,[51] N. Cornish,[101] A. Corsi,[83] S. Cortese,[30] C. A. Costa,[16] M. W. Coughlin,[72,1] S. B. Coughlin,[89] J.-P. Coulon,[67] S. T. Countryman,[51] P. Couvares,[1] P. B. Covas,[102] E. E. Cowan,[77] D. M. Coward,[65] M. J. Cowart,[7] D. C. Coyne,[1] R. Coyne,[83] J. D. E. Creighton,[21] T. D. Creighton,[103] J. Cripe,[2] S. G. Crowder,[104] T. J. Cullen,[29,2] A. Cumming,[46] L. Cunningham,[46] E. Cuoco,[30] T. Dal Canton,[50] G. Dálya,[56] S. L. Danilishin,[22,10] S. D'Antonio,[33] K. Danzmann,[22,10] A. Dasgupta,[105] C. F. Da Silva Costa,[5] V. Dattilo,[30] I. Dave,[61] M. Davier,[28] D. Davis,[44] E. J. Daw,[106] B. Day,[77] S. De,[44] D. DeBra,[52] J. Degallaix,[26] M. De Laurentis,[17,4] S. Deléglise,[71] W. Del Pozzo,[59,23,24] N. Demos,[15] T. Denker,[10] T. Dent,[10] R. De Pietri,[107,108] V. Dergachev,[38] R. De Rosa,[79,4] R. T. DeRosa,[7] C. De Rossi,[26,30] R. DeSalvo,[109] O. de Varona,[10] J. Devenson,[27] S. Dhurandhar,[19] M. C. Díaz,[103] L. Di Fiore,[4] M. Di Giovanni,[110,95] T. Di Girolamo,[51,79,4] A. Di Lieto,[23,24] S. Di Pace,[97,35] I. Di Palma,[97,35] F. Di Renzo,[23,24] Z. Doctor,[91] V. Dolique,[26] F. Donovan,[15] K. L. Dooley,[11] S. Doravari,[10] I. Dorrington,[36] R. Douglas,[46] M. Dovale Álvarez,[59] T. P. Downes,[21] M. Drago,[10] C. Dreissigacker,[10] J. C. Driggers,[47] Z. Du,[82] M. Ducrot,[8] P. Dupej,[46] S. E. Dwyer,[47] T. B. Edo,[106] M. C. Edwards,[72] A. Effler,[7] H.-B. Eggenstein,[38,10] P. Ehrens,[1] J. Eichholz,[1] S. S. Eikenberry,[5] R. A. Eisenstein,[15] R. C. Essick,[15] D. Estevez,[8] Z. B. Etienne,[41] T. Etzel,[1] M. Evans,[15] T. M. Evans,[7] M. Factourovich,[51] V. Fafone,[32,33,17] H. Fair,[44] S. Fairhurst,[36] X. Fan,[82] S. Farinon,[60] B. Farr,[91] W. M. Farr,[59] E. J. Fauchon-Jones,[36] M. Favata,[111] M. Fays,[36] C. Fee,[84] H. Fehrmann,[10] J. Feicht,[1] M. M. Fejer,[52] A. Fernandez-Galiana,[15] I. Ferrante,[23,24] E. C. Ferreira,[16] F. Ferrini,[30] F. Fidecaro,[23,24] D. Finstad,[44] I. Fiori,[30] D. Fiorucci,[39] M. Fishbach,[91] R. P. Fisher,[44] M. Fitz-Axen,[45] R. Flaminio,[26,112] M. Fletcher,[46] H. Fong,[90] J. A. Font,[85,113] P. W. F. Forsyth,[25] S. S. Forsyth,[77] J.-D. Fournier,[67] S. Frasca,[97,35] F. Frasconi,[24] Z. Frei,[56] A. Freise,[59] R. Frey,[70] V. Frey,[28] E. M. Fries,[1] P. Fritschel,[15] V. V. Frolov,[7] P. Fulda,[5] M. Fyffe,[7] H. Gabbard,[46] B. U. Gadre,[19] S. M. Gaebel,[59] J. R. Gair,[114] L. Gammaitoni,[42] M. R. Ganija,[73] S. G. Gaonkar,[19] C. Garcia-Quiros,[102] F. Garufi,[79,4] B. Gateley,[47] S. Gaudio,[37] G. Gaur,[115] V. Gayathri,[116] N. Gehrels,[50,a] G. Gemme,[60] E. Genin,[30] A. Gennai,[24] D. George,[12] J. George,[61] L. Gergely,[117] V. Germain,[8] S. Ghonge,[77] Abhirup Ghosh,[20] Archisman Ghosh,[20,14] S. Ghosh,[66,14,21] J. A. Giaime,[2,7] K. D. Giardina,[7] A. Giazotto,[24] K. Gill,[37] L. Glover,[109] E. Goetz,[118] R. Goetz,[5] S. Gomes,[36] B. Goncharov,[6] G. González,[2] J. M. Gonzalez Castro,[23,24] A. Gopakumar,[119] M. L. Gorodetsky,[62] S. E. Gossan,[1] M. Gosselin,[30] R. Gouaty,[8] A. Grado,[120,4] C. Graef,[46] M. Granata,[26] A. Grant,[46] S. Gras,[15] C. Gray,[47] G. Greco,[121,122] A. C. Green,[59] E. M. Gretarsson,[37] P. Groot,[66] H. Grote,[10] S. Grunewald,[38] P. Gruning,[28] G. M. Guidi,[121,122] X. Guo,[82] A. Gupta,[64] M. K. Gupta,[105] K. E. Gushwa,[1] E. K. Gustafson,[1] R. Gustafson,[118] O. Halim,[18,17] B. R. Hall,[69] E. D. Hall,[15] E. Z. Hamilton,[36] G. Hammond,[46] M. Haney,[123] M. M. Hanke,[10] J. Hanks,[47] C. Hanna,[64] M. D. Hannam,[36] O. A. Hannuksela,[93] J. Hanson,[7] T. Hardwick,[2] J. Harms,[17,18] G. M. Harry,[124] I. W. Harry,[38] M. J. Hart,[46] C.-J. Haster,[90] K. Haughian,[46] J. Healy,[58] A. Heidmann,[71] M. C. Heintze,[7] H. Heitmann,[67] P. Hello,[28] G. Hemming,[30] M. Hendry,[46] I. S. Heng,[46] J. Hennig,[46] A. W. Heptonstall,[1] M. Heurs,[10,22] S. Hild,[46] T. Hinderer,[66] D. Hoak,[30] D. Hofman,[26] K. Holt,[7] D. E. Holz,[91] P. Hopkins,[36] C. Horst,[21] J. Hough,[46] E. A. Houston,[46] E. J. Howell,[65] Y. M. Hu,[10] E. A. Huerta,[12] D. Huet,[28] B. Hughey,[37] S. Husa,[102] S. H. Huttner,[46] T. Huynh-Dinh,[7] N. Indik,[10] R. Inta,[83] G. Intini,[97,35] H. N. Isa,[46] J.-M. Isac,[71] M. Isi,[1] B. R. Iyer,[20] K. Izumi,[47] T. Jacqmin,[71] K. Jani,[77] P. Jaranowski,[125] S. Jawahar,[63] F. Jiménez-Forteza,[102] W. W. Johnson,[2] N. K. Johnson-McDaniel,[13] D. I. Jones,[126] R. Jones,[46] R. J. G. Jonker,[14] L. Ju,[65] J. Junker,[10] C. V. Kalaghatgi,[36] V. Kalogera,[89] B. Kamai,[1] S. Kandhasamy,[7] G. Kang,[40] J. B. Kanner,[1] S. J. Kapadia,[21] S. Karki,[70] K. S. Karvinen,[10] M. Kasprzack,[2] M. Katolik,[12] E. Katsavounidis,[15] W. Katzman,[7] S. Kaufer,[22] K. Kawabe,[47] F. Kéfélian,[67] D. Keitel,[46] A. J. Kemball,[12] R. Kennedy,[106] C. Kent,[36] J. S. Key,[127] F. Y. Khalili,[62] I. Khan,[17,33] S. Khan,[10] Z. Khan,[105] E. A. Khazanov,[128] N. Kijbunchoo,[25] Chunglee Kim,[129] J. C. Kim,[130] K. Kim,[93] W. Kim,[73] W. S. Kim,[131] Y.-M. Kim,[92] S. J. Kimbrell,[77] E. J. King,[73] P. J. King,[47] M. Kinley-Hanlon,[124] R. Kirchhoff,[10] J. S. Kissel,[47] L. Kleybolte,[34] S. Klimenko,[5] T. D. Knowles,[41] P. Koch,[10] S. M. Koehlenbeck,[10] S. Koley,[14] V. Kondrashov,[1] A. Kontos,[15] M. Korobko,[34] W. Z. Korth,[1] I. Kowalska,[74] D. B. Kozak,[1] C. Krämer,[10] V. Kringel,[10] B. Krishnan,[10] A. Królak,[132,133] G. Kuehn,[10] P. Kumar,[90] R. Kumar,[105] S. Kumar,[20] L. Kuo,[87] A. Kutynia,[132] S. Kwang,[21] B. D. Lackey,[38] K. H. Lai,[93] M. Landry,[47] R. N. Lang,[134] J. Lange,[58] B. Lantz,[52] R. K. Lanza,[15] A. Lartaux-Vollard,[28] P. D. Lasky,[6]

M. Laxen,[7] A. Lazzarini,[1] C. Lazzaro,[55] P. Leaci,[97,35] S. Leavey,[46] C. H. Lee,[92] H. K. Lee,[135] H. M. Lee,[136] H. W. Lee,[130] K. Lee,[46] J. Lehmann,[10] A. Lenon,[41] M. Leonardi,[110,95] N. Leroy,[28] N. Letendre,[8] Y. Levin,[6] T. G. F. Li,[93] S. D. Linker,[109] T. B. Littenberg,[137] J. Liu,[65] R. K. L. Lo,[93] N. A. Lockerbie,[63] L. T. London,[36] J. E. Lord,[44] M. Lorenzini,[17,18] V. Loriette,[138] M. Lormand,[7] G. Losurdo,[24] J. D. Lough,[10] C. O. Lousto,[58] G. Lovelace,[29] H. Lück,[22,10] D. Lumaca,[32,33] A. P. Lundgren,[10] R. Lynch,[15] Y. Ma,[48] R. Macas,[36] S. Macfoy,[27] B. Machenschalk,[10] M. MacInnis,[15] D. M. Macleod,[36] I. Magaña Hernandez,[21] F. Magaña-Sandoval,[44] L. Magaña Zertuche,[44] R. M. Magee,[64] E. Majorana,[35] I. Maksimovic,[138] N. Man,[67] V. Mandic,[45] V. Mangano,[46] G. L. Mansell,[25] M. Manske,[21,25] M. Mantovani,[30] F. Marchesoni,[53,43] F. Marion,[8] S. Márka,[51] Z. Márka,[51] C. Markakis,[12] A. S. Markosyan,[52] A. Markowitz,[1] E. Maros,[1] A. Marquina,[100] P. Marsh,[127] F. Martelli,[121,122] L. Martellini,[67] I. W. Martin,[46] R. M. Martin,[111] D. V. Martynov,[15] K. Mason,[15] E. Massera,[106] A. Masserot,[8] T. J. Massinger,[1] M. Masso-Reid,[46] S. Mastrogiovanni,[97,35] A. Matas,[45] F. Matichard,[1,15] L. Matone,[51] N. Mavalvala,[15] N. Mazumder,[69] R. McCarthy,[47] D. E. McClelland,[25] S. McCormick,[7] L. McCuller,[15] S. C. McGuire,[139] G. McIntyre,[1] J. McIver,[1] D. J. McManus,[25] L. McNeill,[6] T. McRae,[25] S. T. McWilliams,[41] D. Meacher,[64] G. D. Meadors,[38,10] M. Mehmet,[10] J. Meidam,[14] E. Mejuto-Villa,[9] A. Melatos,[96] G. Mendell,[47] R. A. Mercer,[21] E. L. Merilh,[47] M. Merzougui,[67] S. Meshkov,[1] C. Messenger,[46] C. Messick,[64] R. Metzdorff,[71] P. M. Meyers,[45] H. Miao,[59] C. Michel,[26] H. Middleton,[59] E. E. Mikhailov,[140] L. Milano,[79,4] A. L. Miller,[5,97,35] B. B. Miller,[89] J. Miller,[15] M. Millhouse,[101] M. C. Milovich-Goff,[109] O. Minazzoli,[67,141] Y. Minenkov,[33] J. Ming,[38] C. Mishra,[142] S. Mitra,[19] V. P. Mitrofanov,[62] G. Mitselmakher,[5] R. Mittleman,[15] D. Moffa,[84] A. Moggi,[24] K. Mogushi,[11] M. Mohan,[30] S. R. P. Mohapatra,[15] M. Montani,[121,122] C. J. Moore,[13] D. Moraru,[47] G. Moreno,[47] S. Morisaki,[81] S. R. Morriss,[103] B. Mours,[8] C. M. Mow-Lowry,[59] G. Mueller,[5] A. W. Muir,[36] A. Mukherjee,[10] D. Mukherjee,[21] S. Mukherjee,[103] N. Mukund,[19] A. Mullavey,[7] J. Munch,[73] E. A. Muñiz,[44] M. Muratore,[37] P. G. Murray,[46] K. Napier,[77] I. Nardecchia,[32,33] L. Naticchioni,[97,35] R. K. Nayak,[143] J. Neilson,[109] G. Nelemans,[66,14] T. J. N. Nelson,[7] M. Nery,[10] A. Neunzert,[118] L. Nevin,[1] J. M. Newport,[124] G. Newton,[46,b] K. K. Y. Ng,[93] T. T. Nguyen,[25] D. Nichols,[66] A. B. Nielsen,[10] S. Nissanke,[66,14] A. Nitz,[10] A. Noack,[10] F. Nocera,[30] D. Nolting,[7] C. North,[36] L. K. Nuttall,[36] J. Oberling,[47] G. D. O'Dea,[109] G. H. Ogin,[144] J. J. Oh,[131] S. H. Oh,[131] F. Ohme,[10] M. A. Okada,[16] M. Oliver,[102] P. Oppermann,[10] R. J. Oram,[7] B. O'Reilly,[7] R. Ormiston,[45] L. F. Ortega,[5] R. O'Shaughnessy,[58] S. Ossokine,[38] D. J. Ottaway,[73] H. Overmier,[7] B. J. Owen,[83] A. E. Pace,[64] J. Page,[137] M. A. Page,[65] A. Pai,[116,145] S. A. Pai,[61] J. R. Palamos,[70] O. Palashov,[128] C. Palomba,[35] A. Pal-Singh,[34] Howard Pan,[87] Huang-Wei Pan,[87] B. Pang,[48] P. T. H. Pang,[93] C. Pankow,[89] F. Pannarale,[36] B. C. Pant,[61] F. Paoletti,[24] A. Paoli,[30] M. A. Papa,[38,21,10] A. Parida,[19] W. Parker,[7] D. Pascucci,[46] A. Pasqualetti,[30] R. Passaquieti,[23,24] D. Passuello,[24] M. Patil,[133] B. Patricelli,[146,24] B. L. Pearlstone,[46] M. Pedraza,[1] R. Pedurand,[26,147] L. Pekowsky,[44] A. Pele,[7] S. Penn,[148] C. J. Perez,[47] A. Perreca,[1,110,95] L. M. Perri,[89] H. P. Pfeiffer,[90,38] M. Phelps,[46] O. J. Piccinni,[97,35] M. Pichot,[67] F. Piergiovanni,[121,122] V. Pierro,[9] G. Pillant,[30] L. Pinard,[26] I. M. Pinto,[9] M. Pirello,[47] M. Pitkin,[46] M. Poe,[21] R. Poggiani,[23,24] P. Popolizio,[30] E. K. Porter,[39] A. Post,[10] J. Powell,[46,149] J. Prasad,[19] J. W. W. Pratt,[37] G. Pratten,[102] V. Predoi,[36] T. Prestegard,[21] M. Prijatelj,[10] M. Principe,[9] S. Privitera,[38] R. Prix,[10] G. A. Prodi,[110,95] L. G. Prokhorov,[62] O. Puncken,[10] M. Punturo,[43] P. Puppo,[35] M. Pürrer,[38] H. Qi,[21] V. Quetschke,[103] E. A. Quintero,[1] R. Quitzow-James,[70] F. J. Raab,[47] D. S. Rabeling,[25] H. Radkins,[47] P. Raffai,[56] S. Raja,[61] C. Rajan,[61] B. Rajbhandari,[83] M. Rakhmanov,[103] K. E. Ramirez,[103] A. Ramos-Buades,[102] P. Rapagnani,[97,35] V. Raymond,[38] M. Razzano,[23,24] J. Read,[29] T. Regimbau,[67] L. Rei,[60] S. Reid,[63] D. H. Reitze,[1,5] W. Ren,[12] S. D. Reyes,[44] F. Ricci,[97,35] P. M. Ricker,[12] S. Rieger,[10] K. Riles,[118] M. Rizzo,[58] N. A. Robertson,[1,46] R. Robie,[46] F. Robinet,[28] A. Rocchi,[33] L. Rolland,[8] J. G. Rollins,[1] V. J. Roma,[70] J. D. Romano,[103] R. Romano,[3,4] C. L. Romel,[47] J. H. Romie,[7] D. Rosińska,[150,57] M. P. Ross,[151] S. Rowan,[46] A. Rüdiger,[10] P. Ruggi,[30] G. Rutins,[27] K. Ryan,[47] S. Sachdev,[1] T. Sadecki,[47] L. Sadeghian,[21] M. Sakellariadou,[152] L. Salconi,[30] M. Saleem,[116] F. Salemi,[10] A. Samajdar,[143] L. Sammut,[6] L. M. Sampson,[89] E. J. Sanchez,[1] L. E. Sanchez,[1] N. Sanchis-Gual,[85] V. Sandberg,[47] J. R. Sanders,[44] B. Sassolas,[26] B. S. Sathyaprakash,[64,36] P. R. Saulson,[44] O. Sauter,[118] R. L. Savage,[47] A. Sawadsky,[34] P. Schale,[70] M. Scheel,[48] J. Scheuer,[89] J. Schmidt,[10] P. Schmidt,[1,66] R. Schnabel,[34] R. M. S. Schofield,[70] A. Schönbeck,[34] E. Schreiber,[10] D. Schuette,[10,22] B. W. Schulte,[10] B. F. Schutz,[36,10] S. G. Schwalbe,[37] J. Scott,[46] S. M. Scott,[25] E. Seidel,[12] D. Sellers,[7] A. S. Sengupta,[153] D. Sentenac,[30] V. Sequino,[32,33,17] A. Sergeev,[128] D. A. Shaddock,[25] T. J. Shaffer,[47] A. A. Shah,[137] M. S. Shahriar,[89] M. B. Shaner,[109] L. Shao,[38] B. Shapiro,[52] P. Shawhan,[76] A. Sheperd,[21] D. H. Shoemaker,[15] D. M. Shoemaker,[77] K. Siellez,[77] X. Siemens,[21] M. Sieniawska,[57] D. Sigg,[47] A. D. Silva,[16] L. P. Singer,[50] A. Singh,[38,10,22] A. Singhal,[17,35] A. M. Sintes,[102] B. J. J. Slagmolen,[25] B. Smith,[7] J. R. Smith,[29] R. J. E. Smith,[1,6] S. Somala,[154] E. J. Son,[131] J. A. Sonnenberg,[21] B. Sorazu,[46] F. Sorrentino,[60] T. Souradeep,[19] A. P. Spencer,[46] A. K. Srivastava,[105] K. Staats,[37] A. Staley,[51] M. Steinke,[10]

J. Steinlechner,[34,46] S. Steinlechner,[34] D. Steinmeyer,[10] S. P. Stevenson,[59,149] R. Stone,[103] D. J. Stops,[59] K. A. Strain,[46] G. Stratta,[121,122] S. E. Strigin,[62] A. Strunk,[47] R. Sturani,[155] A. L. Stuver,[7] T. Z. Summerscales,[156] L. Sun,[96] S. Sunil,[105] J. Suresh,[19] P. J. Sutton,[36] B. L. Swinkels,[30] M. J. Szczepańczyk,[37] M. Tacca,[14] S. C. Tait,[46] C. Talbot,[6] D. Talukder,[70] D. B. Tanner,[5] M. Tápai,[117] A. Taracchini,[38] J. D. Tasson,[72] J. A. Taylor,[137] R. Taylor,[1] S. V. Tewari,[148] T. Theeg,[10] F. Thies,[10] E. G. Thomas,[59] M. Thomas,[7] P. Thomas,[47] K. A. Thorne,[7] E. Thrane,[6] S. Tiwari,[17,95] V. Tiwari,[36] K. V. Tokmakov,[63] K. Toland,[46] M. Tonelli,[23,24] Z. Tornasi,[46] A. Torres-Forné,[85] C. I. Torrie,[1] D. Töyrä,[59] F. Travasso,[30,43] G. Traylor,[7] J. Trinastic,[5] M. C. Tringali,[110,95] L. Trozzo,[157,24] K. W. Tsang,[14] M. Tse,[15] R. Tso,[1] L. Tsukada,[81] D. Tsuna,[81] D. Tuyenbayev,[103] K. Ueno,[21] D. Ugolini,[158] C. S. Unnikrishnan,[119] A. L. Urban,[1] S. A. Usman,[36] H. Vahlbruch,[22] G. Vajente,[1] G. Valdes,[2] M. Vallisneri,[48] N. van Bakel,[14] M. van Beuzekom,[14] J. F. J. van den Brand,[75,14] C. Van Den Broeck,[14] D. C. Vander-Hyde,[44] L. van der Schaaf,[14] J. V. van Heijningen,[14] A. A. van Veggel,[46] M. Vardaro,[54,55] V. Varma,[48] S. Vass,[1] M. Vasúth,[49] A. Vecchio,[59] G. Vedovato,[55] J. Veitch,[46] P. J. Veitch,[73] K. Venkateswara,[151] G. Venugopalan,[1] D. Verkindt,[8] F. Vetrano,[121,122] A. Viceré,[121,122] A. D. Viets,[21] S. Vinciguerra,[59] D. J. Vine,[27] J.-Y. Vinet,[67] S. Vitale,[15] T. Vo,[44] H. Vocca,[42,43] C. Vorvick,[47] S. P. Vyatchanin,[62] A. R. Wade,[1] L. E. Wade,[84] M. Wade,[84] R. Walet,[14] M. Walker,[29] L. Wallace,[1] S. Walsh,[38,10,21] G. Wang,[17,122] H. Wang,[59] J. Z. Wang,[64] W. H. Wang,[103] Y. F. Wang,[93] R. L. Ward,[25] J. Warner,[47] M. Was,[8] J. Watchi,[98] B. Weaver,[47] L.-W. Wei,[10,22] M. Weinert,[10] A. J. Weinstein,[1] R. Weiss,[15] L. Wen,[65] E. K. Wessel,[12] P. Weßels,[10] J. Westerweck,[10] T. Westphal,[10] K. Wette,[25] J. T. Whelan,[58] S. E. Whitcomb,[1] B. F. Whiting,[5] C. Whittle,[6] D. Wilken,[10] D. Williams,[46] R. D. Williams,[1] A. R. Williamson,[66] J. L. Willis,[1,159] B. Willke,[22,10] M. H. Wimmer,[10] W. Winkler,[10] C. C. Wipf,[1] H. Wittel,[10,22] G. Woan,[46] J. Woehler,[10] J. Wofford,[58] K. W. K. Wong,[93] J. Worden,[47] J. L. Wright,[46] D. S. Wu,[10] D. M. Wysocki,[58] S. Xiao,[1] H. Yamamoto,[1] C. C. Yancey,[76] L. Yang,[160] M. J. Yap,[25] M. Yazback,[5] Hang Yu,[15] Haocun Yu,[15] M. Yvert,[8] A. Zadrożny,[132] M. Zanolin,[37] T. Zelenova,[30] J.-P. Zendri,[55] M. Zevin,[89] L. Zhang,[1] M. Zhang,[140] T. Zhang,[46] Y.-H. Zhang,[58] C. Zhao,[65] M. Zhou,[89] Z. Zhou,[89] S. J. Zhu,[38,10] X. J. Zhu,[6] A. B. Zimmerman,[90] M. E. Zucker,[1,15] and J. Zweizig[1]

(LIGO Scientific Collaboration and Virgo Collaboration)

[1]*LIGO, California Institute of Technology, Pasadena, California 91125, USA*
[2]*Louisiana State University, Baton Rouge, Louisiana 70803, USA*
[3]*Università di Salerno, Fisciano, I-84084 Salerno, Italy*
[4]*INFN, Sezione di Napoli, Complesso Universitario di Monte S. Angelo, I-80126 Napoli, Italy*
[5]*University of Florida, Gainesville, Florida 32611, USA*
[6]*OzGrav, School of Physics & Astronomy, Monash University, Clayton 3800, Victoria, Australia*
[7]*LIGO Livingston Observatory, Livingston, Louisiana 70754, USA*
[8]*Laboratoire d'Annecy-le-Vieux de Physique des Particules (LAPP), Université Savoie Mont Blanc, CNRS/IN2P3, F-74941 Annecy, France*
[9]*University of Sannio at Benevento, I-82100 Benevento, Italy and INFN, Sezione di Napoli, I-80100 Napoli, Italy*
[10]*Albert-Einstein-Institut, Max-Planck-Institut für Gravitationsphysik, D-30167 Hannover, Germany*
[11]*The University of Mississippi, University, Mississippi 38677, USA*
[12]*NCSA, University of Illinois at Urbana-Champaign, Urbana, Illinois 61801, USA*
[13]*University of Cambridge, Cambridge CB2 1TN, United Kingdom*
[14]*Nikhef, Science Park, 1098 XG Amsterdam, Netherlands*
[15]*LIGO, Massachusetts Institute of Technology, Cambridge, Massachusetts 02139, USA*
[16]*Instituto Nacional de Pesquisas Espaciais, 12227-010 São José dos Campos, São Paulo, Brazil*
[17]*Gran Sasso Science Institute (GSSI), I-67100 L'Aquila, Italy*
[18]*INFN, Laboratori Nazionali del Gran Sasso, I-67100 Assergi, Italy*
[19]*Inter-University Centre for Astronomy and Astrophysics, Pune 411007, India*
[20]*International Centre for Theoretical Sciences, Tata Institute of Fundamental Research, Bengaluru 560089, India*
[21]*University of Wisconsin-Milwaukee, Milwaukee, Wisconsin 53201, USA*
[22]*Leibniz Universität Hannover, D-30167 Hannover, Germany*
[23]*Università di Pisa, I-56127 Pisa, Italy*
[24]*INFN, Sezione di Pisa, I-56127 Pisa, Italy*
[25]*OzGrav, Australian National University, Canberra, Australian Capital Territory 0200, Australia*
[26]*Laboratoire des Matériaux Avancés (LMA), CNRS/IN2P3, F-69622 Villeurbanne, France*
[27]*SUPA, University of the West of Scotland, Paisley PA1 2BE, United Kingdom*
[28]*LAL, Univ. Paris-Sud, CNRS/IN2P3, Université Paris-Saclay, F-91898 Orsay, France*

[29]California State University Fullerton, Fullerton, California 92831, USA
[30]European Gravitational Observatory (EGO), I-56021 Cascina, Pisa, Italy
[31]Chennai Mathematical Institute, Chennai 603103, India
[32]Università di Roma Tor Vergata, I-00133 Roma, Italy
[33]INFN, Sezione di Roma Tor Vergata, I-00133 Roma, Italy
[34]Universität Hamburg, D-22761 Hamburg, Germany
[35]INFN, Sezione di Roma, I-00185 Roma, Italy
[36]Cardiff University, Cardiff CF24 3AA, United Kingdom
[37]Embry-Riddle Aeronautical University, Prescott, Arizona 86301, USA
[38]Albert-Einstein-Institut, Max-Planck-Institut für Gravitations-physik, D-14476 Potsdam-Golm, Germany
[39]APC, AstroParticule et Cosmologie, Université Paris Diderot, CNRS/IN2P3, CEA/Irfu, Observatoire de Paris, Sorbonne Paris Cité, F-75205 Paris Cedex 13, France
[40]Korea Institute of Science and Technology Information, Daejeon 34141, Korea
[41]West Virginia University, Morgantown, West Virginia 26506, USA
[42]Università di Perugia, I-06123 Perugia, Italy
[43]INFN, Sezione di Perugia, I-06123 Perugia, Italy
[44]Syracuse University, Syracuse, New York 13244, USA
[45]University of Minnesota, Minneapolis, Minnesota 55455, USA
[46]SUPA, University of Glasgow, Glasgow G12 8QQ, United Kingdom
[47]LIGO Hanford Observatory, Richland, Washington 99352, USA
[48]Caltech CaRT, Pasadena, California 91125, USA
[49]Wigner RCP, RMKI, H-1121 Budapest, Konkoly Thege Miklós út 29-33, Hungary
[50]NASA Goddard Space Flight Center, Greenbelt, Maryland 20771, USA
[51]Columbia University, New York, New York 10027, USA
[52]Stanford University, Stanford, California 94305, USA
[53]Università di Camerino, Dipartimento di Fisica, I-62032 Camerino, Italy
[54]Università di Padova, Dipartimento di Fisica e Astronomia, I-35131 Padova, Italy
[55]INFN, Sezione di Padova, I-35131 Padova, Italy
[56]Institute of Physics, Eötvös University, Pázmány P. s. 1/A, Budapest 1117, Hungary
[57]Nicolaus Copernicus Astronomical Center, Polish Academy of Sciences, 00-716, Warsaw, Poland
[58]Rochester Institute of Technology, Rochester, New York 14623, USA
[59]University of Birmingham, Birmingham B15 2TT, United Kingdom
[60]INFN, Sezione di Genova, I-16146 Genova, Italy
[61]RRCAT, Indore MP 452013, India
[62]Faculty of Physics, Lomonosov Moscow State University, Moscow 119991, Russia
[63]SUPA, University of Strathclyde, Glasgow G1 1XQ, United Kingdom
[64]The Pennsylvania State University, University Park, Pennsylvania 16802, USA
[65]OzGrav, University of Western Australia, Crawley, Western Australia 6009, Australia
[66]Department of Astrophysics/IMAPP, Radboud University Nijmegen, P.O. Box 9010, 6500 GL Nijmegen, Netherlands
[67]Artemis, Université Côte d'Azur, Observatoire Côte d'Azur, CNRS, CS 34229, F-06304 Nice Cedex 4, France
[68]Institut FOTON, CNRS, Université de Rennes 1, F-35042 Rennes, France
[69]Washington State University, Pullman, Washington 99164, USA
[70]University of Oregon, Eugene, Oregon 97403, USA
[71]Laboratoire Kastler Brossel, UPMC-Sorbonne Universités, CNRS, ENS-PSL Research University, Collège de France, F-75005 Paris, France
[72]Carleton College, Northfield, Minnesota 55057, USA
[73]OzGrav, University of Adelaide, Adelaide, South Australia 5005, Australia
[74]Astronomical Observatory Warsaw University, 00-478 Warsaw, Poland
[75]VU University Amsterdam, 1081 HV Amsterdam, Netherlands
[76]University of Maryland, College Park, Maryland 20742, USA
[77]Center for Relativistic Astrophysics, Georgia Institute of Technology, Atlanta, Georgia 30332, USA
[78]Université Claude Bernard Lyon 1, F-69622 Villeurbanne, France
[79]Università di Napoli "Federico II", Complesso Universitario di Monte S.Angelo, I-80126 Napoli, Italy
[80]Dipartimento di Fisica, Università degli Studi di Genova, I-16146 Genova, Italy
[81]RESCEU, University of Tokyo, Tokyo, 113-0033, Japan
[82]Tsinghua University, Beijing 100084, China
[83]Texas Tech University, Lubbock, Texas 79409, USA
[84]Kenyon College, Gambier, Ohio 43022, USA
[85]Departamento de Astronomía y Astrofísica, Universitat de València, E-46100 Burjassot, València, Spain
[86]Museo Storico della Fisica e Centro Studi e Ricerche Enrico Fermi, I-00184 Roma, Italy

[87]National Tsing Hua University, Hsinchu City, 30013 Taiwan, Republic of China
[88]Charles Sturt University, Wagga Wagga, New South Wales 2678, Australia
[89]Center for Interdisciplinary Exploration & Research in Astrophysics (CIERA), Northwestern University, Evanston, Illinois 60208, USA
[90]Canadian Institute for Theoretical Astrophysics, University of Toronto, Toronto, Ontario M5S 3H8, Canada
[91]University of Chicago, Chicago, Illinois 60637, USA
[92]Pusan National University, Busan 46241, Korea
[93]The Chinese University of Hong Kong, Shatin, NT, Hong Kong
[94]INAF, Osservatorio Astronomico di Padova, I-35122 Padova, Italy
[95]INFN, Trento Institute for Fundamental Physics and Applications, I-38123 Povo, Trento, Italy
[96]OzGrav, University of Melbourne, Parkville, Victoria 3010, Australia
[97]Università di Roma "La Sapienza", I-00185 Roma, Italy
[98]Université Libre de Bruxelles, Brussels 1050, Belgium
[99]Sonoma State University, Rohnert Park, California 94928, USA
[100]Departamento de Matemáticas, Universitat de València, E-46100 Burjassot, València, Spain
[101]Montana State University, Bozeman, Montana 59717, USA
[102]Universitat de les Illes Balears, IAC3–IEEC, E-07122 Palma de Mallorca, Spain
[103]The University of Texas Rio Grande Valley, Brownsville, Texas 78520, USA
[104]Bellevue College, Bellevue, Washington 98007, USA
[105]Institute for Plasma Research, Bhat, Gandhinagar 382428, India
[106]The University of Sheffield, Sheffield S10 2TN, United Kingdom
[107]Dipartimento di Scienze Matematiche, Fisiche e Informatiche, Università di Parma, I-43124 Parma, Italy
[108]INFN, Sezione di Milano Bicocca, Gruppo Collegato di Parma, I-43124 Parma, Italy
[109]California State University, Los Angeles, 5151 State University Drive, Los Angeles, California 90032, USA
[110]Università di Trento, Dipartimento di Fisica, I-38123 Povo, Trento, Italy
[111]Montclair State University, Montclair, New Jersey 07043, USA
[112]National Astronomical Observatory of Japan, 2-21-1 Osawa, Mitaka, Tokyo 181-8588, Japan
[113]Observatori Astronòmic, Universitat de València, E-46980 Paterna, València, Spain
[114]School of Mathematics, University of Edinburgh, Edinburgh EH9 3FD, United Kingdom
[115]University and Institute of Advanced Research, Koba Institutional Area, Gandhinagar Gujarat 382007, India
[116]IISER-TVM, CET Campus, Trivandrum Kerala 695016, India
[117]University of Szeged, Dóm tér 9, Szeged 6720, Hungary
[118]University of Michigan, Ann Arbor, Michigan 48109, USA
[119]Tata Institute of Fundamental Research, Mumbai 400005, India
[120]INAF, Osservatorio Astronomico di Capodimonte, I-80131, Napoli, Italy
[121]Università degli Studi di Urbino "Carlo Bo", I-61029 Urbino, Italy
[122]INFN, Sezione di Firenze, I-50019 Sesto Fiorentino, Firenze, Italy
[123]Physik-Institut, University of Zurich, Winterthurerstrasse 190, 8057 Zurich, Switzerland
[124]American University, Washington, DC 20016, USA
[125]University of Biał ystok, 15-424 Biał ystok, Poland
[126]University of Southampton, Southampton SO17 1BJ, United Kingdom
[127]University of Washington Bothell, 18115 Campus Way NE, Bothell, Washington 98011, USA
[128]Institute of Applied Physics, Nizhny Novgorod, 603950, Russia
[129]Korea Astronomy and Space Science Institute, Daejeon 34055, Korea
[130]Inje University Gimhae, South Gyeongsang 50834, Korea
[131]National Institute for Mathematical Sciences, Daejeon 34047, Korea
[132]NCBJ, 05-400 Świerk-Otwock, Poland
[133]Institute of Mathematics, Polish Academy of Sciences, 00656 Warsaw, Poland
[134]Hillsdale College, Hillsdale, Michigan 49242, USA
[135]Hanyang University, Seoul 04763, Korea
[136]Seoul National University, Seoul 08826, Korea
[137]NASA Marshall Space Flight Center, Huntsville, Alabama 35811, USA
[138]ESPCI, CNRS, F-75005 Paris, France
[139]Southern University and A&M College, Baton Rouge, Louisiana 70813, USA
[140]College of William and Mary, Williamsburg, Virginia 23187, USA
[141]Centre Scientifique de Monaco, 8 quai Antoine Ier, MC-98000, Monaco
[142]Indian Institute of Technology Madras, Chennai 600036, India
[143]IISER-Kolkata, Mohanpur, West Bengal 741252, India
[144]Whitman College, 345 Boyer Avenue, Walla Walla, Washington 99362, USA
[145]Indian Institute of Technology Bombay, Powai, Mumbai, Maharashtra 400076, India

[146]*Scuola Normale Superiore, Piazza dei Cavalieri 7, I-56126 Pisa, Italy*
[147]*Université de Lyon, F-69361 Lyon, France*
[148]*Hobart and William Smith Colleges, Geneva, New York 14456, USA*
[149]*OzGrav, Swinburne University of Technology, Hawthorn VIC 3122, Australia*
[150]*Janusz Gil Institute of Astronomy, University of Zielona Góra, 65-265 Zielona Góra, Poland*
[151]*University of Washington, Seattle, Washington 98195, USA*
[152]*King's College London, University of London, London WC2R 2LS, United Kingdom*
[153]*Indian Institute of Technology, Gandhinagar Ahmedabad Gujarat 382424, India*
[154]*Indian Institute of Technology Hyderabad, Sangareddy, Khandi, Telangana 502285, India*
[155]*International Institute of Physics, Universidade Federal do Rio Grande do Norte, Natal RN 59078-970, Brazil*
[156]*Andrews University, Berrien Springs, Michigan 49104, USA*
[157]*Università di Siena, I-53100 Siena, Italy*
[158]*Trinity University, San Antonio, Texas 78212, USA*
[159]*Abilene Christian University, Abilene, Texas 79699, USA*
[160]*Colorado State University, Fort Collins, Colorado 80523, USA*

[a]Deceased, February 2017.
[b]Deceased, December 2016.